# Colloidal Micromotors: Controlled Directed Motion

Larysa Baraban[a], Christian Kreidler, Denys Makarov, Paul Leiderer, and Artur Erbe

Department of Physics, University of Konstanz, Universitätstrasse 10, Konstanz, D-78457, Germany

Here we demonstrate a synthetic micro-engine, based on long-range controlled movement of colloidal particles, which is induced by a local catalytic reaction. The directed motion at long timescales was achieved by placing specially designed magnetic capped colloids in a hydrogen peroxide solution at weak magnetic fields. The control of the motion of the particles was provided by changes of the concentration of the solution and by varying the strength of the applied magnetic field. Such synthetic objects can then be used not only to understand the fundamental driving processes but also be employed as small motors in biological environments, for example, for the transportation of molecules in a controllable way.

82.70.Dd, 82.39.-k

---

[a] To whom correspondence should be addressed; e-mail: Larysa.Baraban@uni-konstanz.de

Chemical processes, which make life of cells and bacteria possible, rely on the existence of directed transport of certain substances to relevant places. On the other hand, the behaviour of microscopic objects is governed by random Brownian motion. Biological engines combine random and directed contributions of motion. From this interplay of energy scales and the microscopic nature of the building blocks an exceptional complexity of the cellular driving mechanisms arises. The common functional principle of biological engines is the conversion of chemical energy of the environment into mechanical work [1, 5]. Prominent examples are cytoplasmic molecular motors [1−3] – the proteins, which use ATP hydrolysis [4] to perform movement. The complexity of all such structures often impedes fundamental understanding of the underlying processes. In contrast to molecular driving processes [1−5], artificial motors reveal simple working principles. Recently colloidal particles were suggested as appropriate objects for the development of synthetic nano- and micro-engines [6, 7]. However, studies of colloidal transport in fluids showed the necessity of an external *macroscopic* influence (*e.g.* gradient fields [8−11]) in order to generate the movement of particles. Consequently, in this approach a macroscopic response of a colloidal ensemble is achieved, without control of the motion of individual particles. In contrast to the previously mentioned *macroscopic* approach, another concept was applied, where every particle is driven by its own *local* motive force, excluding the necessity of the external influence [12, 13]. Although the directed motion at short timescales was successfully achieved, the control of motion at long timescales was not realized [13].

Here we demonstrate the concept of a combined *local-macroscopic* influence on colloidal particles, which explores the transition from the regime of Brownian random walk to fully directed motion. This is achieved by using silica spheres (diameter $4.75\,\mu m$) with artificially designed magnetic moments. Magnetic properties were provided by deposition of a multilayer stack $Pt(1\,nm)/[Co(0.3\,nm)/Pt(0.8\,nm)]_8/Pt(5\,nm)$ with out-of-plane magnetic anisotropy onto densely packed monolayer of particles, as it is described previously [14–16]. For the purpose of the current experiment, the sample after deposition was magnetically saturated in order to induce a non-zero out-of-plane remanent magnetic moment $\boldsymbol{m}$, directed along the axis of the capped particles (Fig. 1 a). Finally, a top layer of 25 nm Pd was evaporated as a catalyst for the chemical reaction. After fabrication the capped colloids were detached from the surface [16] and suspended in an aqueous solution of $H_2O_2$ [17]. The suspension was placed into the measurement cell [14], where particles are settled on the glass substrate due to gravity. In the following we study the motion of the capped particles on the glass surface in the $xy$-plane (Fig. 1 b). Being mixed with solution of $H_2O_2$ these particles experience a driving force due to the chemical decomposition of hydrogen peroxide at the surface of the cap, which is catalyzed by Pd. The balance equation $2\,H_2O_2 \rightarrow 2\,H_2O + O_2$ gives the ratio of reactants to products of 2:3. Consequently a local gradient of the chemical potential around the particle is formed, resulting in the appearance of a difference in osmotic pressure $\Delta p$ (Fig. 1 b). This imposes propulsion of the sphere from its Pd side (high pressure region) towards the uncoated silica part (low pressure region). A similar method to induce motion of macroscopic objects by chemical reaction was first suggested by Ismagilov *et al.* [18]. The recent experimental study by Howse *et al.* [13] revealed the possibility to obtain the directed motion of microscopic beads at short timescales. In this letter we demonstrate that application of an external magnetic field in $y$-direction orient magnetic caps along the field lines, allowing to obtain directed motion of the particles at long timescales. It thus opens the possibility to use them as micro-vehicles for transportation of synthetic and biological objects.

The movement of the particles in our experiment was imaged by means of video microscopy [16]. In the following the mean squared displacement (MSD) is used to characterize the motion of capped particles in solutions with different concentrations of fuel-molecules $n(H_2O_2)$ in an external magnetic field $B$, applied along $y$-axis. In the absence of an external magnetic field and fuel-molecules ($B = 0\,\text{mT}$; $n(H_2O_2) = 0\%$) the particles demonstrate a classical Brownian behavior, which is proven by a linear increase of the MSD with time (Fig. 2 a, inset I a). The displacement of the particle is governed by processes of translational and rotational diffusion with coefficients $D_T = k_B T / 6\pi\eta R_0$ and $D_R = k_B T / 8\pi\eta R_0^3$, respectively. Here, $k_B T$ is the thermal energy, $\eta$ is the viscosity of the solution, and $R_0$ is the radius of the particle. Once fuel-molecules are added, the local chemical reaction causes the propulsion of the sphere. It leads to an enhancement of the fluctuations of the particles owing to the increase of the translational component of the diffusion in solution, as it is demonstrated in Fig. 2 a, inset II a for $n(H_2O_2) = 30\%$. The analytical expression for MSD of a particle propelling with velocity $v$, which couples the contributions from rotational and translational diffusions was evaluated by Howse *et al.* [13]. Limiting forms of this equation are:

$$t \ll \tau_R: \quad \Delta r^2(t) = 4D_T t + v t^2 \tag{1a}$$

$$t \gg \tau_R: \quad \Delta r^2(t) = D^* t \equiv (4D_T + v\tau_R)t \tag{1b},$$

where $\tau_R = 8\pi\eta R_0^3 / k_B T$ is the rotational diffusion time. Thus, the transition from directed motion, characterized by parabolic MSD at $t \ll \tau_R$ to Brownian-like motion with an apparently larger diffusion constant $D^*$ (linear MSD with time) at long timescales is predicted by Eq. 1 (Fig. 2 a). The randomization of motion is imposed by processes of rotational diffusion with the characteristic timescale $\tau_R$, which can be estimated from the inflection point of the curves in Fig. 2 a. We found a nonlinear monotonic decrease of $\tau_R$ with concentration of $H_2O_2$: from 70 s at $n(H_2O_2) = 3\%$ to 30 s at $n(H_2O_2) = 30\%$, which is in agreement with results reported by Howse *et al.* [13]. The reduction of the rotational diffusion time might be caused by the higher probability to enhance angular fluctuations of the bead due to local inhomogeneities on the surface of caps or/and in $H_2O_2$ concentrations near the particle surface.

As the randomization of motion in $H_2O_2$ solution is caused by rotational diffusion, the suppression of this process should allow directed motion of particles at long timescales. Due to the presence of the magnetic moment, the capped particle can be aligned in a *homogeneous* magnetic field, which imposes a restriction on the rotational degree of freedom. The trajectories (Fig. 2 b, insets) show that the application of a weak magnetic field $B = 0.2$ mT along *y*-axis leads to the directed long-scale propulsion of the particle. In contrast to the low-scale fluctuations in *x*-direction, the displacement in *y*-direction reveals rectilinear movement with MSD quadratically dependent on time (Fig. 2 b). In order to demonstrate the possibility of manipulation the catalytically propelled beads, the direction of the applied magnetic field was successively changed from being parallel or antiparallel to the *y*-axis reversing the direction of the particle motion. The corresponding trajectory is presented in Fig. 2 b, inset II b.

For various applications, like transportation of molecules or biological cells, the control of the motion parameters, i.e. direction and speed of transfer, is important. The velocity of the particles along the field direction $v_y$ depends on the concentration of the fuel-molecules (Fig. 3 a). For instance, in solutions with $n(H_2O_2) = 20\%$ and $B = 0.2\,mT$ the velocity of the bead is $v_y = 6\,\mu m/s$. The velocity stays proportional to the concentration of $H_2O_2$ up to $n(H_2O_2) = 20\%$, gradually approaching saturation at its higher values. This is caused by the specific kinetics of the chemical reaction and is discussed by Robbins *et al.* [19].

Whereas the dependence of the velocity of particles on the concentration of the fuel-molecules is to be expected, it is not obvious that the magnetic field should have any influence other than orient a particle. The observation of a distinct magnetic field dependence of the velocity $v_y(B)$ as depicted in Fig. 3 b was therefore even more surprising. At low magnetic field the velocity of the particle is governed by the catalytic reaction. However, an increase of the field up to about $2\,mT$ leads to slowing down the bead until it stops completely. A further increase of the magnetic field causes a reversal of the direction of motion toward the palladium side of the particle. We would like to emphasize, that this effect is not provoked by an *inhomogeneity* of the applied magnetic field. Test measurements carried out in pure water did not reveal any additional displacement of the particles caused by magnetic field, as it would be expected in the case of a present magnetic field gradient. Therefore, we can claim that the competing driving mechanism, which reverses the motion of particles, has another origin. In addition, this reversal of the direction of motion is found to be strongly dependent on the concentration of $H_2O_2$ (Fig. 3 b). We therefore assume that a driving counter-mechanism has to be related to the catalytic reaction as well as to the applied magnetic field. The following simple model is proposed in order to get a qualitative understanding of the underlying physics. The decomposition of $H_2O_2$ is accompanied with the formation of ions, i.e. $OH^-$, $H^+$, etc. [20]. The recombination of ions after a time $\tau_0$ leads to the creation of $H_2O$ and $O_2$ resulting in the appearance of the osmotic force. The presence of ions near the Pd side of the colloid induces a gradient electric field $E$. This electric field polarizes the particle and gives rise to a dipole force $F_{dip} = \alpha \cdot \nabla E^2$, with $\alpha$ – polarizability of the colloid [21]. Similar to laser trapping [21, 22], this force attracts a dielectric particle into the region of the highest electric field gradient resulting in the particle's motion *towards* its palladium side. Since the dipole force is caused by ions at the Pd side of the particle, the resulting velocity of the colloids has to depend on the life-time of ions. In an applied magnetic field trajectories of ions are modified according to the Lorentz force. In order to evaluate these changes, we use the well-known expression for the mean-free path of the ion $\lambda = (\sigma\, n_i)^{-1}$, where the scattering cross-section $\sigma$ is modified accounting the Lorentz force: $\sigma = \pi R^2 \sim B^{-2}$, leading to the magnetic field dependence of life-time of the ion; $n_i$ is the volume concentration of ions [23]. For a rough estimation we assume, that the electric potential has Debye form $\varphi = \dfrac{\Theta \cdot x_D}{\varepsilon_0} \exp(-x/x_D)$, where $\Theta$ is the area charge density on the surface of the particle and $x_D$ is the Debye screening length [24]. This allows to obtain a simple expression for the resulting velocity of the particle in a magnetic field: $v_y(B) = v_{osm} - v_{dip} \cdot (B/B_0)^2$, where $v_{osm}$ is the contribution given by the osmotic force; the second term is derived from the dipole force: $v_{dip} = \dfrac{2\alpha\Theta^2}{\varepsilon_0^2 M_P} \cdot \dfrac{\tau_0}{x_D}$, $M_P$ is the mass of the colloid, $B_0$ is a fitting constant related to the modification of the life-time of ions. Consequently, with an increase of the magnetic field the particle slows down and at certain critical field $B_C = B_0 \cdot \sqrt{v_{osm}/v_{dip}}$ it reverses direction of its motion. At larger values of the magnetic field ($B > B_C$) the propulsion of the particle is determined mainly by the dipole force. Using typical values for the parameters [25], we estimate $v_{dip}$ to be about 1 μm/s, which by order of magnitude agrees with experimentally measured value. Thus, this dipole force *might* be considered as a counter-mechanism discussed above. However, in order to get a better understanding of this intriguing phenomenon, more detailed theoretical considerations are still required.

In conclusion, a controlled directed motion of capped colloidal particles with artificially designed magnetic properties is demonstrated. The propulsion of the beads is induced by an asymmetric catalytic reaction. The application of a weak *homogeneous* magnetic field blocks processes of rotational diffusion and thus allows to achieve directed motion of the particle at long-scales. In addition, it is observed that a magnetic field determines not only the direction of the motion of the particle, but also influences its velocity. This combined effect allows us to control parameters of the motion of the particle *in situ* by means of slight changes of the external magnetic field. Such fully controlled microscopic engines can be used, for example, to transport biological objects and molecules, attached to the functionalized surface of the particles.

## ACKNOWLEDGMENTS

We would like to thank Prof. Dr. M. Albrecht (Chemnitz University of Technology) for experimental help. The work was supported by International Research Training Group "Soft Condenced Matter".

**Figure captions**

**FIG. 1** Schematic description of the system. (a) Capped particle covered by a Pd / [Co / Pt]$_8$ multilayer stack. The direction of the easy axis of magnetization is sketched. (b) Geometry of the experiment: the directed motion of a capped particle on a glass substrate is induced by a catalytic chemical reaction in an applied in-plane magnetic field. A *homogeneous* magnetic field $B$ is applied along the $y$-axis.

**FIG. 2** Propulsion of colloidal particles: from Brownian random walk to directed motion. MSD of the beads in solutions with different concentrations of H$_2$O$_2$: (a) without applied magnetic field. $\Delta r^2(t)$ reveals a parabolic behaviour at short timescales $t << \tau_R$ and switches to linear regime at $t >> \tau_R$. This statement is supported by fitting the experimental data (symbols) with parabolic and linear functions (lines), respectively. The trajectories of the particle are demonstrated in insets I a and II a; (b) in a weak applied magnetic field, $B = 0.2$ mT. MSD in this case depends quadratically on time, which is peculiar for rectilinear motion. Corresponding trajectories of the propelling beads are given in insets I b and II b. The parabolic fit (lines) of experimental data (symbols) is provided.

**FIG. 3** Manipulation of the catalytically propelled particles. (a) The concentration dependence of the velocity of a directed motion of the particles, $v_y$, in an applied magnetic field, $B = 0.2$ mT. (b) Dependencies of $v_y$ on the strength of the applied magnetic field $B$.

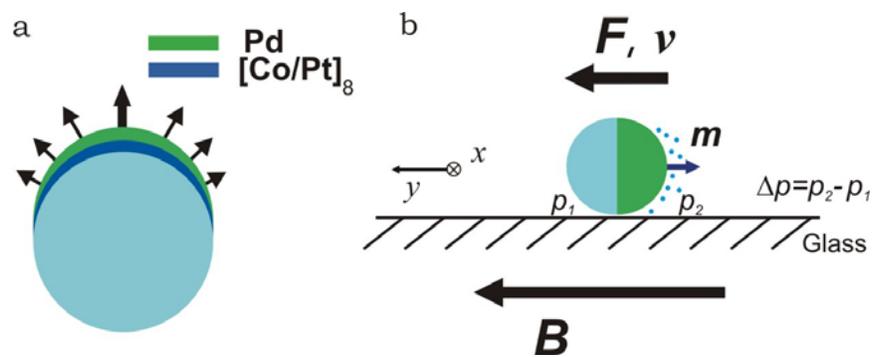

**FIG. 1**

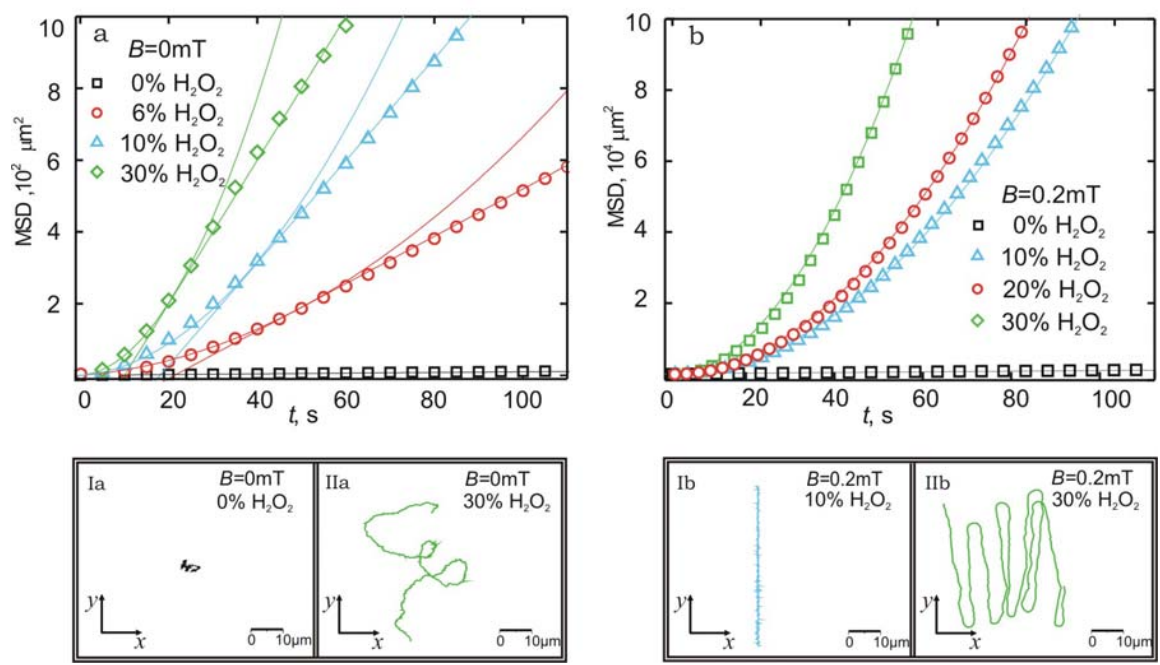

**FIG. 2**

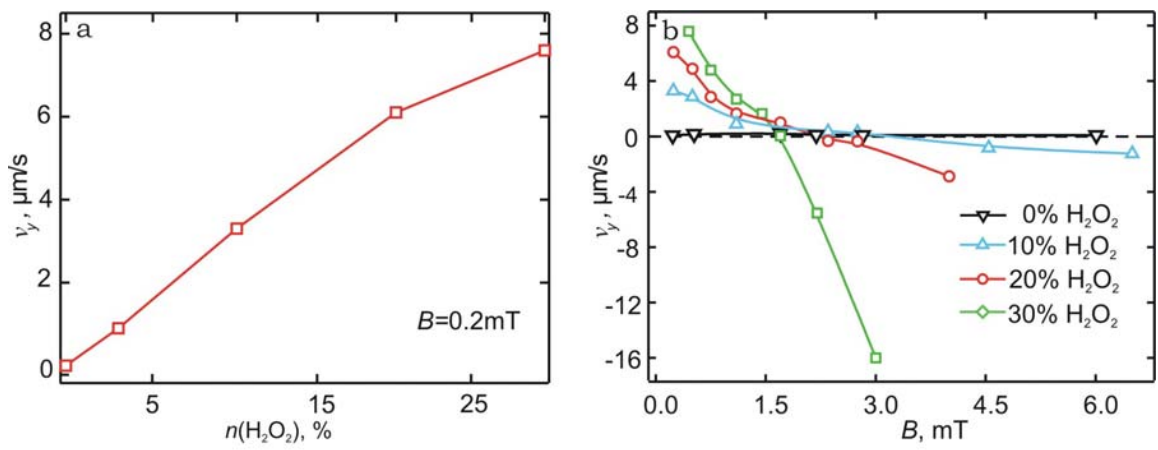

**FIG. 3**